\documentclass{article}

\usepackage{arxiv}

\usepackage[utf8]{inputenc} 
\usepackage[T1]{fontenc}    
\usepackage{url}            
\usepackage{booktabs}       
\usepackage{amsfonts}       
\usepackage{nicefrac}       
\usepackage{microtype}      
\usepackage{lipsum}
\usepackage{subfigure}
\usepackage{amsmath}
\usepackage{mathtools}
\usepackage[T1]{fontenc}
\usepackage{hyperref}
\usepackage{subfigure}

\title{Multiscale Phase Retrieval}

\author{
David J. Brady\\
Wyant College of Optical Sciences\\ University of Arizona\\
Tucson, Az 85721 \\
  \texttt{djbrady@arizona.edu}\\
  \And
Timothy J. Schulz\\
Department of Electrical and Computer Engineering\\ Michigan Technological University\\  Houghton, MI 49931\\
\texttt{schulz@mtu.edu}\\
\And
Chengyu Wang\\
Department of Electrical and Computer Engineering\\
Duke University\\
Durham, NC 27705 \\
\texttt{chengyu.wang@duke.edu}\\
}

\begin{document}
\maketitle

\begin{abstract}
While characterization of coherent wavefields is essential to laser, x-ray and electron imaging, sensors measure the squared magnitude of the field, rather than the field itself. Holography or phase retrieval must be used to characterize the field. The need for a reference severely restricts the utility of holography. Phase retrieval, in contrast, is theoretically consistent with sensors that directly measure coherent or partially coherent fields with no prior assumptions. Unfortunately, phase retrieval has not yet been successfully implemented for large-scale fields. Here we show that both holography and phase retrieval are capable of quantum-limited coherent signal estimation and we describe phase retrieval strategies that approach the quantum limit  for >1 megapixel fields. These strategies rely on group testing using networks of interferometers, such as might be constructed using emerging integrated photonic, plasmonic and/or metamaterial devices. Phase-sensitive sensor planes using such devices could eliminate the need both for lenses and reference signals, creating a path to large aperture diffraction limited laser imaging. 
\end{abstract}

\section*{Introduction}
Characterization of a coherent wave field under the constraint that only the squared modulus of the wave is detectable has been addressed by holography~\cite{gabor1948new, schnars2015digital} and by phase retrieval~\cite{Gerchberg72, Gonsalves:76, fienup1982phase}. Holography utilizes a reference signal---something Gabor called a "coherent background." When the reference is sufficiently intense, the measured intensities can be linearly processed to recover the field. In contrast, without the benefit of a reference signal, phase-retrieval requires estimation of the field from a quadratic form~\cite{chen2017solving}. Here we utilize interconnected arrays of low rank transformations to reduce computational complexity and enable practical reference-free measurement of megapixel fields. We specifically show that $5 n\log n$ appropriately coded measurements are sufficient to estimate the phase and amplitude of megapixel scale signals to within 10\% of the quantum limited mean square error (MSE). 

Phase retrieval, under various names and methods, has been explored in many different contexts. Classically defined "phase retrieval" as pioneered in the above cited studies by Gerchberg, Saxton, Fienup and Gonslaves began with the challenge of reconstructing a signal from the modulus of its Fourier transform, but has recently expanded to include analysis of more general quadratic forms~\cite{chen2017solving}. From another perspective, wavefront estimation from shearing interferometry~\cite{southwell1980wave, Rimmer} has long been used in adaptive optics and astronomy. Phase estimation also lies at the center of two mode interferometry, as explored extensively in gravity wave detection and quantum state analysis~\cite{caves1981quantum}. This paper applies algorithms from conventional phase retrieval to physical measurements related to shearing and two mode interferometry. The advantage of this approach is that low rank measurements reduce the physical and computational complexity of phase retrieval while enabling near quantum limited field estimation. The main contributions of this paper are:
\begin{enumerate}
    \item to show that the quantum limited lower bound on the mean square error for estimating a coherent field consisting of $N$ modes is $N$;
    \item to show that phase retrieval can achieve this bound; and
    \item to describe and simulate practical phase retrieval measurements that approach this bound.
\end{enumerate}
While we expect that improved results may be achieved by applying signal priors and regularization, such methods are not necessary for the results presented here and are thus reserved for future analysis. Similarly, we anticipate that improved results might be obtained using prepared quantum states, but here we limit our attention to semiclassical analysis using a  Poisson noise model. 

Phase retrieval consists of estimating the complex field $\mathbf{x}$ from the magnitude or intensity of the linearly transformed field  $\mathbf{y}=\mathbf{A}\mathbf{x}$~\cite{shechtman2015phase}. While early studies focused on $\mathbf{A}$ as a Fourier kernel, recent studies have modified the forward model by adding coding elements in the Fourier plane~\cite{wang2017ultra,horisaki2014single, candes2015phase}. Fourier coding is closely related to multiband filtering, which is called "Fourier ptychography"~\cite{rodenburg2007hard}. Expanding this approach, numerous recent mathematical studies have considered phase retrieval under the assumption that $\mathbf{A}$ is a random matrix.
Estimating $\mathbf{x}$ given the measurements $|y_m|^2$ and the matrix $\mathbf{A}$ is known to be NP-hard \cite{wang2017solving} and is typically solved by iterative algorithms. While the diversity of algorithms is large, novel algorithms have not dramatically exceeded the signal fidelity of classic iterative approaches~\cite{chandra2017phasepack}. However, recent studies have led to better understanding of the role of the coding matrix $\mathbf{A}$. Design of measurement matrices has a long history in linear measurement systems, as evidenced by Hadamard spectroscopy~\cite{harwit1979hadamard} and uniformly redundant arrays~\cite{fenimore1978coded}, but to date has been little considered for quadratic forms. 

 Shearing interferometry~\cite{strojnik2007lateral} and wavefront sensors~\cite{malacara2007hartmann} can also be used for phase retrieval. The matrix $\mathbf{A}$ is related to the gradient operator in this case, and numerous algorithms have been developed to estimate the field from measurement of the gradient~\cite{gilles2002multigrid,talmi2006wavefront,mochi2015modal}. Unfortunately, integration of the gradient is not well conditioned. One can improve conditioning by making the local sampling more sophisticated~\cite{tumbar2000wave}. One can view this approach as subgroup testing using nearest neighbors, here we show that the use nonlocal groups further improves estimation.   
 
 Phase estimation strategies for two mode interferometry have been explored by many fundamental and applied studies in the context of gravity wave detection and quantum state analysis~\cite{caves1981quantum}. Most such studies do not consider estimation of the relative amplitudes of the modes. With respect to phase estimation, the "shot noise limit" yields an estimator deviation inversely proportional to $\sqrt{\bar{K}}$, where $\bar{K}$ is the expected number of photons in the phase measurement. Using entangled states of light, it is possible to improve on shot noise to reach the "Heisenberg" limit, where the deviation is inversely proportional to $\bar{K}$~\cite{caves1981quantum, pezze2015phase}. The success of phase measurement in two mode systems may be viewed as an exitence proof that quantum limited phase retrieval is possible. Here we use such systems to scale to large scale fields. 
 
We seek to combine insights from Fourier phase-retrieval, wavefront sensing and two-mode phase estimation in the design of measurement matrix $\mathbf{A}$. We are inspired in this regard by more than a half century of work considering holographic wavefront reconstruction as a coding and communication problem~\cite{leith1962reconstructed}. Our approach focuses on "multiscale" analysis, under which networks of low rank interferometers are interconnected to reconstruct high dimensional fields. We are able to prove that as the dimension rank of the interferometer and the field increase this approach asymptotically reaches quantum limited performance. We then show in simulation that this approach yields error  within 10\% of the quantum limit for more practical measurements. We then further extend the multiscale concept by simulating block-based phase retrieval, which linearizes the computational cost per pixel at the cost of increasing MSE by a constant factor. 

Focal imaging systems rely on the concept of a "focal plane array," e.g. a generic sensor that can digitize an image field. An analogous device for coherent field characterization is the vision of this paper. In imagining such a device, we must consider what would be analogous to a the focal plane's pixel, e.g. what is the atomic component of the measurement. We propose that joint measurement of small subsets of the unknown field provides this atomic kernel. We find below that arrays of 2 to 5 mode interferometers can be used to scale phase retrieval to arbitrarily large fields. Appropriate arrays processing a modest number of modes have been implemented via integrated optics\cite{Carolan711,Clements:16}. Use of such systems for interferometric imaging is described by Su, {\it et al.}~\cite{su2018interferometric}. We hope that the mathematical results presented here will spur further development of such systems. Among many potential applications, such sensor arrays might enable lensless imaging over large apertures without the need for a reference beam. This vision is contained in the idea of "holographic aperture ladar,"~\cite{Marron:92, Stafford:16} with the exception that here we suggest that the "coherent background" of holographic---or {\em whole field} --- measurement is not necessary. 

\section*{Measurements and Fisher Information}
This section addresses the general problem of estimating a coherent field from the intensities of an affine transformation of the field. In doing so, we arrive at four significant results: 1) a lower bound for the squared estimation-error that any system can attain; 2) a necessary and sufficient condition to determine if a measurement system induces this optimal bound; 3) a simple proof that phase-quadrature holography is optimal; and 4) an existence proof that phase-retrieval is asymptotically optimal. 

A continuous coherent field $g(u)$ in an aperture plane can be described by a discrete vector of modal amplitudes $\mathbf{x}$ by projection onto the basis of orthonormal mode distributions $\{ \psi_n(u) \}$ according to 
\begin{equation}
g(u) = \sum_{n=1}^N x_n \psi_n(u).
\end{equation}
The intensity of the field is $\| \mathbf{x} \|^2$, and---when using the semiclassical theory for detection---this intensity specifies the expected number of photo-detections we associate with the field: 
\begin{align}
\bar{K} 
& = \| \mathbf{x} \|^2 \nonumber \\
& = \sum_{n=1}^N \left|x_n\right|^2.
\end{align}
Characterization of the vector of mode amplitudes, $\mathbf{x}$, requires intensity measurements that encode both its amplitude and phase. Holography and phase retrieval accomplish this by measuring the intensity of an affine transformation of the modes:
\begin{equation}
\mathbf{y} = {\boldsymbol\rho} + \mathbf{A} \mathbf{x},
\label{eq:measurementTransformation}
\end{equation}
where $\mathbf{A}$ is an $M \times N$ orthonormal matrix with $M > N$, and ${\boldsymbol\rho}$ is an $M$-element non-negative reference signal.  Orthonormality of $\mathbf{A}$ ensures that the measurements conserve the field's intensity. 

The intensities $|y_m|^2$ may be detected consistent with the semiclassical theory for detection with 
\begin{equation}
\Pr\left[ \mathbf{d} = \mathbf{k} | \mathbf{x} \right]
= \prod_{m=1}^M
e^{-  \left| y_m \right|^2 }
\left( \left| y_m \right|^2 \right)^{k_m}
\frac{1}{k_m!}.
\label{eq:detectionModel}
\end{equation}
We assume that the $m$th row of the measurement matrix $\mathbf{A}$ is $\mathbf{a}_m^\dagger$, where we use the superscript $^\dagger$ to denote the conjugate transpose of a matrix or vector, and the superscript $^T$ to denote the standard transpose. 
If we parameterize the complex-valued mode vector by its real and imaginary parts:
\begin{equation}
\mathbf{x} = \mathbf{r} + j \mathbf{i},
\end{equation}
then the squared error for an estimate $\widehat{\mathbf{x}}$ is
\begin{align}
{\cal E}^2
& = \sum_{n=1}^N \left| \widehat{x}_n - x_n \right|^2 \nonumber \\
& = \sum_{n=1}^N \left( \widehat{r}_n - r_n \right)^2 + 
\sum_{n=1}^N \left( \widehat{i}_n - i_n \right)^2 ,
\end{align}
and the conditional mean-square error is:
\begin{align}
\text{MSE}(\mathbf{x})
& = E\left[ {\cal E}^2 | \mathbf{x} \right] \nonumber \\
& = \sum_{n=1}^N 
E\left[ \left| \widehat{x}_n - x_n \right|^2 | \mathbf{x} \right] \nonumber \\
& = \sum_{n=1}^N 
E\left[ \left( \widehat{r}_n - r_n \right)^2 | \mathbf{x} \right] + 
\sum_{n=1}^N
E\left[ \left( \widehat{i}_n - i_n \right)^2 | \mathbf{x} \right].
\end{align}
Whereas this mean-square error is, in general, different for each estimator, we can use the Fisher information and the associated Cram\'{e}r-Rao lower bound\cite{kay:1993} to place a lower bound on the mean-square error for any unbiased estimator. 

\subsection*{Cram\'{e}r-Rao Lower Bound}
The Cram\'{e}r-Rao lower bound on the mean-square error for the estimation of the mode vector $\mathbf{x} = \mathbf{r} + j\mathbf{i}$ is:
\begin{equation}
\text{MSE}(\mathbf{x})
\geq
\text{Trace}\left\{ \mathbf{J}^{-1} \right\}
\end{equation}
where the matrix $\mathbf{J}$ is the Fisher information matrix for the estimation of the parameters $\mathbf{r}$ and $\mathbf{i}$ based on the statistical model for the measurements in Eq.~(\ref{eq:detectionModel}). Because quantum-limited measurements of the intensities are Poisson distributed, the Fisher information matrix is \cite{snyder:miller:1991}:
\begin{equation}
\mathbf{J} = 
\left[ 
\begin{array}{cc}
\mathbf{J}_{r,r} & \mathbf{J}_{r,i} \\
\mathbf{J}_{i,r} & \mathbf{J}_{i,i}
\end{array}
\right],
\end{equation}
where
\begin{equation}
\mathbf{J}_{\beta,\zeta} = 
\sum_{m = 1}^M
\frac{\partial  \left| y_m \right|^2}{\partial {\boldsymbol\beta}}
\frac{\partial  \left| y_m \right|^2}{\partial {\boldsymbol\zeta}^T} 
\frac{1}{\left| y_m \right|^2},
\end{equation}
with $\boldsymbol\beta$ and $\boldsymbol\zeta$ taking all possible combinations of $\mathbf{r}$ and $\mathbf{i}$. 
Because 
\begin{equation}
\frac{\partial  \left| y_m \right|^2}{\partial \mathbf{r}}
= \mathbf{a}_m^* y_m^* + \mathbf{a}_m y_m ,
\end{equation}
and
\begin{equation}
\frac{\partial  \left| y_m \right|^2}{\partial \mathbf{i}}
=  j \mathbf{a}_m^* y_m^*  - j \mathbf{a}_m y_m,
\end{equation}
the blocks of the Fisher information are:
\begin{equation}
\mathbf{J}_{r,r} = 2 \mathbf{I} + \mathbf{C}_R,
\end{equation}
\begin{equation}
\mathbf{J}_{i,i} = 2 \mathbf{I} - \mathbf{C}_R,
\end{equation}
and
\begin{equation}
\mathbf{J}_{r,i} = \mathbf{J}_{i,r} = \mathbf{C}_I,
\end{equation}
where
\begin{align}
\mathbf{C}_R 
& = 2 \mbox{Re} \left\{
\sum_{m=1}^M \mathbf{a}_m \mathbf{a}_m^T \frac{ y_m^2  }{\left| y_m \right|^2} \right\},
\end{align}
\begin{equation}
\mathbf{C}_I = 2 \mbox{Im} \left\{
\sum_{m=1}^M \mathbf{a}_m \mathbf{a}_m^T  \frac{ y_m^2 }{\left| y_m \right|^2} \right\},
\end{equation}
and the orthonormality of $\mathbf{A}$ results in:
\begin{align}
\sum_{m=1}^M \mathbf{a}_m \mathbf{a}_m^\dagger 
& = \mathbf{A}^\dagger \mathbf{A} \nonumber \\
& = \mathbf{I}.
\end{align}
Accordingly, the Fisher information matrix is
\begin{equation}
\mathbf{J} = 2 \mathbf{I} + 
\left[
\begin{array}{cc}
\mathbf{C}_R & \mathbf{C}_I \\
\mathbf{C}_I & - \mathbf{C}_R 
\end{array}
\right],
\end{equation}
and its inverse satisfies:
\begin{equation}
\mbox{Trace} \left\{\mathbf{J}^{-1} \right\} \geq N, 
\end{equation}
with equality when $\mathbf{C}_R = \mathbf{C}_I = \mathbf{0}$, or, equivalently:
\begin{align}
\mathbf{C}
& = \mathbf{C}_R + j \mathbf{C}_I \nonumber \\
& = 
\sum_{m=1}^M \mathbf{a}_m \mathbf{a}_m^T  \frac{ y_m^2 }{\left| y_m \right|^2} \nonumber \\
& = \mathbf{0}.
\end{align}
To see this, we note that all the eigenvalues of the $2N \times 2N$ symmetric matrix  
\begin{equation}
\mathbf{C}_{RI} = 
\left[
\begin{array}{cc}
\mathbf{C}_R & \mathbf{C}_I \\
\mathbf{C}_I & - \mathbf{C}_R 
\end{array}
\right],
\end{equation}
must be greater than $-2$ because the eigenvalues of the Fisher information matrix $\mathbf{J}$ must be greater than $0$. Furthermore, if $\gamma$ is an eigenvalue for $\mathbf{C}_{RI}$, then $-\gamma$ is also an eigenvalue because 
\begin{equation}
\left[
\begin{array}{cc}
\mathbf{C}_R & \mathbf{C}_I \\
\mathbf{C}_I & - \mathbf{C}_R 
\end{array}
\right]
\left[
\begin{array}{c}
\mathbf{u}_1 \\ \mathbf{u}_2
\end{array}
\right]
= 
\gamma 
\left[
\begin{array}{c}
\mathbf{u}_1 \\ \mathbf{u}_2
\end{array}
\right],
\end{equation}
implies that
\begin{equation}
\left[
\begin{array}{cc}
\mathbf{C}_R & \mathbf{C}_I \\
\mathbf{C}_I & - \mathbf{C}_R 
\end{array}
\right]
\left[
\begin{array}{c}
-\mathbf{u}_2 \\ \mathbf{u}_1
\end{array}
\right]
= 
-\gamma 
\left[
\begin{array}{c}
-\mathbf{u}_2 \\ \mathbf{u}_1
\end{array}
\right].
\end{equation}
This means the $2N$ eigenvalues for the Fisher information are $\lambda_n = 2 \pm \gamma_n$, and, consequently, the trace for the inverse of this matrix is
\begin{align}
\label{eq:crlb}
\mbox{Tr}\left\{ \mbox{CRLB} \right\}
& = 
\sum_{n=1}^N \left( 
\frac{1}{2 + \gamma_n}  + 
\frac{1}{2 - \gamma_n} 
\right) \nonumber \\
& = 
\sum_{n=1}^N 
\frac{4}{4 - \gamma_n^2} \nonumber \\
& \geq N,
\end{align}
with equality when all the eigenvalues for $\mathbf{C}_{RI}$ are equal to zero, or when $\mathbf{C}_R = \mathbf{C}_I = \mathbf{0}$. 

\subsection*{Holographic Measurement}
One way to accomplish this condition is to make the reference signal large ($\rho_m \gg \max\{\mathbf{x}\}$) and select the matrix $\mathbf{A}$ so that $\mathbf{A}^T \mathbf{A} = \mathbf{0}$. Suppose, for example, the matrix $\mathbf{A}$ directs each mode to four measurements, and each measurement contains only one mode. That is, $\mathbf{A}$ is a $4N \times N$ matrix:
\begin{equation}
\mathbf{A} = 
\left[
\begin{array}{cccc}
\mathbf{a} & & & \\
& \mathbf{a} & & \\
& & \ddots & \\
& & & \mathbf{a}
\end{array}
\right],
\end{equation}
with 
\begin{equation}
\mathbf{a} = 
\frac{1}{\sqrt{4}} \left[
\begin{array}{c}
1 \\ j \\ -1 \\ -j
\end{array}
\right].
\end{equation}
This corresponds to phase quadrature holography~\cite{lai2000wave,Liu:11}, and, when the reference amplitudes $\{ \rho_m \}$ are sufficiently large, we have 
\begin{equation}
\frac{ y_m^2 }{\left| y_m \right|^2} \rightarrow 1,
\end{equation}
and
\begin{align}
\sum_{m=1}^M \mathbf{a}_m \mathbf{a}_m^T  \frac{ y_m^2 }{\left| y_m \right|^2} 
& \rightarrow \mathbf{A}^T \mathbf{A} \nonumber \\
& = \mathbf{0},
\end{align}
so the system attains the quantum limit for field estimation. 

\subsection*{Phase Retrieval}
When we use a reference signal for the field estimation, all the phases for the modes $\mathbf{x}$ are defined relative to the phase for the reference. Without a reference signal, we typically use the phase for one of the modes as our reference phase. To accomplish this, we can arbitrarily define the imaginary part for some mode to be zero, and reduce the estimation from $2N$ parameters to $2N-1$ parameters. 

Even without the benefit of a reference signal, we can still get arbitrarily close to the quantum-limit for the estimation error when we have a large number of modes $N$. This result is achieved by using a group-testing strategy. In contrast with holography---where we take 4 measurements of each mode---for phase retrieval we take $Q$ measurements of each group of $L\geq 2$ modes. To demonstrate that phase retrieval measurements can induce the optimal bound, we assume we make measurements on every such group. Because the total number of groups is $T = \binom{N}{L}$, the overall measurement matrix takes the form
\begin{equation}
\mathbf{A}
= 
\frac{1}{\sqrt{K}}
\left[ 
\begin{array}{c}
\mathbf{W} \mathbf{S}_1 \\
\mathbf{W} \mathbf{S}_2 \\
\vdots \\
\mathbf{W} \mathbf{S}_T 
\end{array}
\right]
\end{equation}
where $\mathbf{W}$ is a $Q \times L$ matrix with $Q \geq L$. We ensure that $\mathbf{W}$, and by extension $\mathbf{A}$, are  orthonormal by setting
\begin{equation}
\label{eq:wql}
W_{q,l} = \frac{1}{\sqrt{Q}} e^{j \frac{2\pi}{Q} q l},
\end{equation}
and $\mathbf{S}_t$ as an $L \times N$ binary matrix
that selects $L$-element combinations from the mode vector $\mathbf{x}$. For example, if $N=4$ and $L=2$, the matrix $\mathbf{S}$ that selects modes $1$ and $3$ is 
\begin{equation}
\mathbf{S}
= 
\left[ 
\begin{array}{cccc}
1&0&0&0 \\
0&0&1&0 \\
\end{array}
\right]
\end{equation}
The factor $1/\sqrt{K}$ accounts for the fact that each mode splits its intensity across $K = \binom{N-1}{L-1}$ of the $T$ combinations. Whereas each mode interferes with itself in $Q K$ of these measurements, it interferes with each of the other modes in $QP$ of the measurements, where $P = \binom{N-2}{L-2}$ so the ratio of self- to cross-interference is
\begin{equation}
\frac{P}{K} = \frac{L-1}{N-1}.
\end{equation}
Now, because the elements of $\mathbf{A}$ are such that
\begin{equation}
|a_{m,n}| = \frac{1}{\sqrt{QK}}, 
\end{equation}
the magnitude for
\begin{equation}
C_{n,r} = 
\sum_{m=1}^M a_{m,n} a_{m,r}  \frac{ y_m^2 }{\left| y_m \right|^2},
\end{equation}
satisfies:
\begin{align}
\left| C_{n,r}  \right| 
& = \left| \sum_{m=1}^M a_{m,n} a_{m,r}  \frac{ y_m^2 }{\left| y_m \right|^2} \right| \nonumber \\
& \leq \sum_{m=1}^M \left| a_{m,n} \right| \left| a_{m,r} \right|  \nonumber \\
& = \frac{1}{QK} \sum_{m=1}^M \alpha_{m,n} \alpha_{m,r} \nonumber \\
& = 
\left\{
\begin{array}{cc}
1 & n = r \\
\displaystyle \frac{L-1}{N-1} & n \neq r
\end{array}
\right. ,
\end{align}
where $\alpha_{m,n}$ is equal to $1$ if $x_n$ contributes to the $m$th measurement, and equal to $0$ otherwise. 
For fixed $L$, then, the matrix $\mathbf{C}$ becomes diagonal as $N \rightarrow \infty$, and the trace of the $(2N-1) \times (2N-1)$ Cram\'{e}r-Rao bound matrix is
\begin{equation}
\mbox{Trace} \left\{ \mathbf{J}^{-1} \right\} = 
\frac{1}{2 + u_1} 
+ \sum_{n=2}^{N} \frac{4}{4 - u_n^2 - v_n^2},
\end{equation}
where 
\begin{equation}
u_n = 2 \mbox{Re}\left\{ C_{n,n}  \right\},
\end{equation}
\begin{equation}
v_n = 2 \mbox{Im}\left\{ C_{n,n} \right\},
\end{equation}
and, to account for the overall phase, we've assumed that the imaginary part for $x_1$ is zero. 

To analyze the performance bound, then, we can consider 
\begin{align}
|C_{n,n}|^2 
& = 
\sum_{m=1}^M \sum_{p=1}^M 
a_{m,n}^2 
(a_{p,n}^*)^2 
\frac{ y_m^2 }{\left| y_m \right|^2} 
\frac{ (y_p^*)^2 }{\left| y_p \right|^2}  \nonumber \\
& = \sum_{m=1}^M \sum_{p=1}^M 
a_{m,n}^2 
(a_{p,n}^*)^2 
e^{j2(\theta_m - \theta_p)},
\end{align}
where $\theta_m$ is the phase for $y_m$. $|C_{n,n}|^2$ will, in general, depend on both the measurement matrix $\mathbf{A}$ and the values for the modes $\mathbf{x}$. Suppose, though, we model the uncertainty about $\mathbf{x}$ as a zero-mean random vector with an average intensity per mode equal to $N_p$:
\begin{equation}
E\left[ \frac{1}{N} \sum_{n=1}^N |x_n|^2 \right] = N_p.
\end{equation}
The maximum-entropy distribution for this constraint is a zero-mean, complex Gaussian random vector with diagonal covariance 
\begin{equation}
E\left[ \mathbf{x} \mathbf{x}^\dagger \right]
= N_p \mathbf{I},
\end{equation}
and, in this situation, $\mathbf{y}$ is a zero-mean, complex Gaussian random vector with the covariance:
\begin{align}
E\left[ \mathbf{y} \mathbf{y}^\dagger \right]
& = \mathbf{R} \nonumber \\
& = N_p \mathbf{A} \mathbf{A}^\dagger.
\end{align}
Then, if we denote 
\begin{equation}
e^{j2\theta_m} = 
\frac{y_m^2}{|y_m|^2},
\end{equation}
this phase signal will be zero-mean with the covariance \cite{miller:1974}:
\begin{align}
E[ e^{j2(\theta_m - \theta_p)} ]
& = \Gamma_{m,p} \nonumber \\
& = 
e^{j2 \phi_{m,p}} \left[ 1 + \frac{(1-\rho_{m,p}^2) \ln (1 - \rho_{m,p}^2)}{\rho_{m,p}^2} \right],
\end{align}
where $\phi_{m,p}$ is the phase and  $\rho_{m,p}^2$ is the amplitude of
\[
\frac{R^2_{m,p}}{R_{m,m} R_{p,p}}.
\]
Because 
\begin{equation}
\left[ 1 + \frac{(1-\rho_{m,p}^2) \ln (1 - \rho_{m,p}^2)}{\rho_{m,p}^2} \right] \leq \rho_{m,p}^2,
\end{equation}
we have that
\begin{align}
E \left[ |C_{n,n}|^2 \right]
& = 
\sum_{m=1}^M \sum_{p=1}^M
a_{m,n}^2 (a_{p,n}^*)^2 
\Gamma_{m,p} \nonumber \\
& \leq  \frac{1}{(QK)^2} \sum_{m=1}^M \sum_{p=1}^M 
\alpha_{m,n} \alpha_{p,n}
 \rho_{m,p}^2.
\end{align}
We've utilized the inequality here to form an upper bound on the lower bound. If, instead, we wanted to analyze the bound for a particular value of $L$, we could use the exact form for this expectation. 
 
Now, the form for our measurement matrix $\mathbf{A}$ is such that
\begin{align}
\mathbf{R}
& = \mathbf{A} \mathbf{A}^\dagger \nonumber \\
& = 
\frac{1}{K}
\left[
\begin{array}{cccc}
\mathbf{W} \mathbf{W}^\dagger & 
\mathbf{W} \mathbf{S}_1 \mathbf{S}_2^T \mathbf{W}^\dagger & 
\cdots & 
\mathbf{W} \mathbf{S}_1 \mathbf{S}_T^T \mathbf{W}^\dagger  \\
\mathbf{W} \mathbf{S}_2 \mathbf{S}_1^T \mathbf{W}^\dagger & 
\mathbf{W} \mathbf{W}^\dagger & 
\cdots &
\mathbf{W} \mathbf{S}_2 \mathbf{S}_T^T \mathbf{W}^\dagger  \\
\vdots & & \ddots & \vdots \\
\mathbf{W} \mathbf{S}_T \mathbf{S}_1^T \mathbf{W}^\dagger  &
\mathbf{W} \mathbf{S}_T \mathbf{S}_2^T \mathbf{W}^\dagger  &
\cdots & 
\mathbf{W} \mathbf{W}^\dagger 
\end{array}
\right],
\end{align}
where $\mathbf{S}_t \mathbf{S}_t^T = \mathbf{I}$, and 
\begin{align}
\left[ \mathbf{W} \mathbf{W}^\dagger \right]_{q,r}
& = \frac{1}{Q}
\sum_{l=1}^{L-1} e^{j\frac{2\pi}{Q} l(q-r) } \nonumber \\
& = \frac{1}{Q} e^{j \frac{\pi (L-1)}{Q} (q-r)} \frac{\sin \pi \frac{L}{Q}(q-r)}{\sin \pi \frac{1}{Q} (q-r)},
\end{align}
so that the diagonal elements of $\mathbf{R}$ are
\begin{equation}
R_{m,m} = \frac{L}{KQ}.
\end{equation}

To evaluate the upper bound for $E[|C_{n,n}|^2]$, we need to evaluate $\rho_{m,p}^2$ for the $K^2$ blocks of $\mathbf{R}$ that have a contribution from $x_n$, which includes $K$ diagonal blocks and $K(K-1)$ off-diagonal blocks. To do this, we'll need to normalize each block by $(\frac{L}{KQ})^2$, then sum the squared magnitudes for those normalized values. 

To better understand each of the off-diagonal blocks for some value for $n$, suppose that $n = 2$, $L = 5$, and the combination corresponding to $\mathbf{S}_1$ contains $x_1, x_2, x_3, x_8$ and $x_{9}$, and the combination corresponding to $\mathbf{S}_5$ contains $x_2, x_4, x_7, x_9$ and $x_{10}$. The two combinations share the modes $x_2$ and $x_9$, so that
\begin{align}
\frac{1}{K}
\left[ \mathbf{W} \mathbf{S}_1 \mathbf{S}_5^T \mathbf{W}^\dagger\right]_{q,r} 
& = 
\frac{1}{K}
\left[ 
\mathbf{w}_2 \mathbf{w}_1^\dagger + 
\mathbf{w}_5 \mathbf{w}_4^\dagger
\right]_{q,r} \nonumber \\
& = \frac{1}{QK} 
\left[ 
e^{j\frac{2\pi}{Q} 2q} e^{-j\frac{2\pi}{Q} 1r} + 
e^{j\frac{2\pi}{Q} 5q} e^{-j\frac{2\pi}{Q} 4r} 
\right], 
\end{align}
where $\mathbf{w}_l$ is the $l$th column of $\mathbf{W}$. We get this result because $x_2$ is in the second position in the first combination, first position in the second combination; and $x_9$ is in the fifth position in the first combination, fourth position in the second combination. Now, because
\begin{align}
\left| 
\frac{1}{K}
\left[ \mathbf{W} \mathbf{S}_1 \mathbf{S}_5^T \mathbf{W}^\dagger\right]_{q,r} 
\right|^2
& = 
\left( \frac{1}{QK} \right)^2
\left| 
e^{j\frac{2\pi}{Q} 2q} e^{-j\frac{2\pi}{Q} 1r} + 
e^{j\frac{2\pi}{Q} 5q} e^{-j\frac{2\pi}{Q} 4r} 
\right|^2 \nonumber \\
& \leq \left( \frac{1}{QK} \right)^2 2^2, 
\end{align}
we can infer the general result that
\begin{equation}
\left| 
\frac{1}{K}
\left[ \mathbf{W} \mathbf{S}_t \mathbf{S}_{\tau}^T \mathbf{W}^\dagger\right]_{q,r} 
\right|^2
\leq
 \left( \frac{1}{QK} \right)^2 L_{t,\tau}^2,
\end{equation}
where $L_{t,\tau}$ is the number of modes in common between the $t$th and $\tau$th combinations. If we divide by $( \frac{L}{KQ} )^2$ to obtain the corresponding element of $\rho^2$, then sum over each block, we obtain the result that the contribution from each block that contains $x_n$ to the sum  that defines $E[|C_{n,n}|^2]$  is less than or equal to $Q^2 L^2_{t,\tau}/L^2$. The total contribution for the $K^2$ blocks is then
\begin{align}
\sum_{m=1}^M \sum_{p=1}^M \alpha_{m,n} \alpha_{p,n} \rho^2_{m,p} 
\leq \frac{Q^2}{L^2} \sum_{(t,\tau) \in {\cal T}_n } L^2_{t,\tau},
\end{align}
where ${\cal T}_n$ is the $K^2$ pairs of $L$-element combinations $(t, \tau)$ that include $x_n$. 

The number of modes in common between the $t$th and $\tau$th combinations, $L_{t,\tau}$, must be less than or equal to $L$ and greater than or equal to $1$. When $N$ is much greater than $L$, though, nearly all of the values for $L_{t,\tau}$ are equal to one. This follows from the fact that the probability of no common elements in any two randomly-selected combinations of $L-1$ elements from $N-1$ is equal to 
\begin{equation}
\frac{\displaystyle \binom{N-L}{L-1}}{\displaystyle \binom{N-1}{L-1}}\approx \frac{\left (1-\frac{L}{N}\right )^{2N-2L}}{ \left (1-\frac{2L}{N}\right )^{N-2L}}\xrightarrow{N \gg L} 1. 
\end{equation}
where we use Stirling's approximation to expand the factorials. 
In the limit of $N \gg L$, then, we can place an upper bound on $E[|C_{n,n}|^2]$:
\begin{align}
E \left[ |C_{n,n}|^2 \right]
& = 
\sum_{m=1}^M \sum_{p=1}^M
a_{m,n}^2 (a_{p,n}^*)^2 
\Gamma_{m,p} \nonumber \\
& \leq  \frac{1}{(QK)^2} \sum_{m=1}^M \sum_{p=1}^M 
\alpha_{m,n} \alpha_{p,n}
 \rho_{m,p}^2 \nonumber \\
 & \leq \frac{1}{L^2}.
\end{align}
Therefore, by selecting $L$ large enough, a phase-retrieval system can get arbitrarily close to the quantum-limit for field estimation in the limit of a large number of modes. Of course when $N$ and $L$ are large, the number of combinations $T$ is generally impractical, but this result shows that a phase-retrieval system can encode all the information needed to attain the quantum-limit and achieve a mean-square error equal to the total number of modes, or a per mode error equal to 1.

\section*{Example Measurement Systems}

This section presents simulations of phase retrieval systems derived from the strategy of the previous section. We consider cases from $L=2$ to $L=5$. Recognizing that measurement of $\binom{N}{L}$ groups as suggested in the last section is impractical, we instead consider randomly selected subsets of the groups proposed above. Here we show that $L n\log{}n$ measurements are sufficient to approach the quantum limit and that such measurements allow computationally tractable phase retrieval. As suggested in the previous section, increasing $L$ reduces MSE.

For $L=2$, we assume random connections which require that 1) each mode in {\bf x} is connected pairwise to $p= 2\log_2 n$ randomly selected modes, and 2) there does not exist a subset $\bf \tilde{x}\subset x$ such that no connections exist between modes in $\bf \tilde{x}$ and modes in $\bf x \backslash  \tilde{x}$. This is easily understood as a connected graph with $n$ vertices, and each vertex has $p$ adjacent vertices. Each group of modes is measured using $\mathbf{W}$ as defined in Eqn.~\ref{eq:wql} with $Q=3$. In total, we consider $n\log_2 n$ pairs and $3 n\log_2 n$ measurements. For $L=3,4,5$ we select $Q=L$. The connection strategy for $L>2$ is the same as for $L = 2$, each connection randomly draws $L$ modes. For $n = 2^k$, each mode is measured/selected $L\log n$ times. The number of total measurements is $Ln\log n$.

\begin{figure}[htbp]
  \centering
  \subfigure[MSE]{
    \label{fig:random_l_mse} 
    \includegraphics[width=0.47\textwidth]{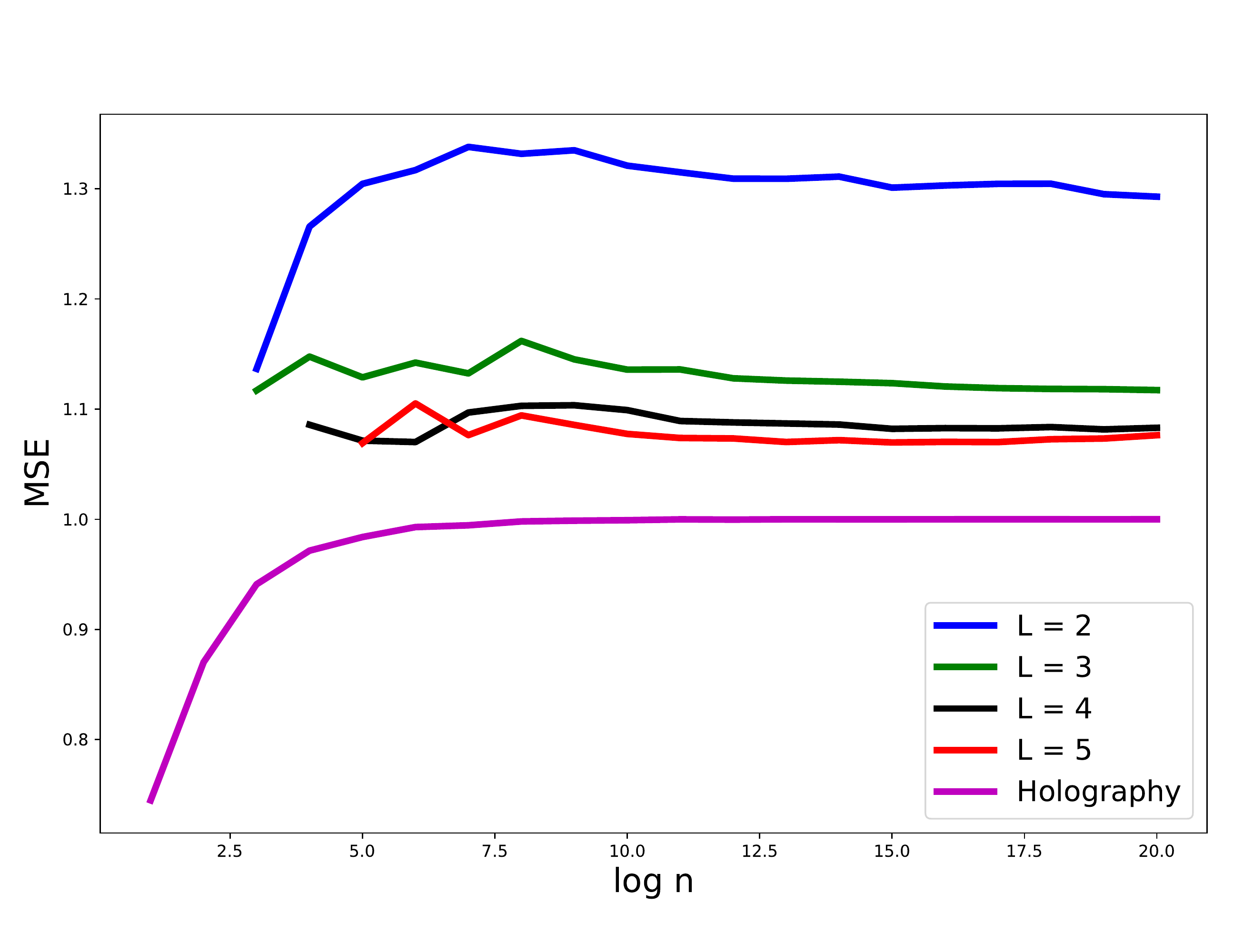}}
  \hspace{0.1in}
  \subfigure[TIME]{
    \label{fig:random_l_time} 
    \includegraphics[width=0.47\textwidth]{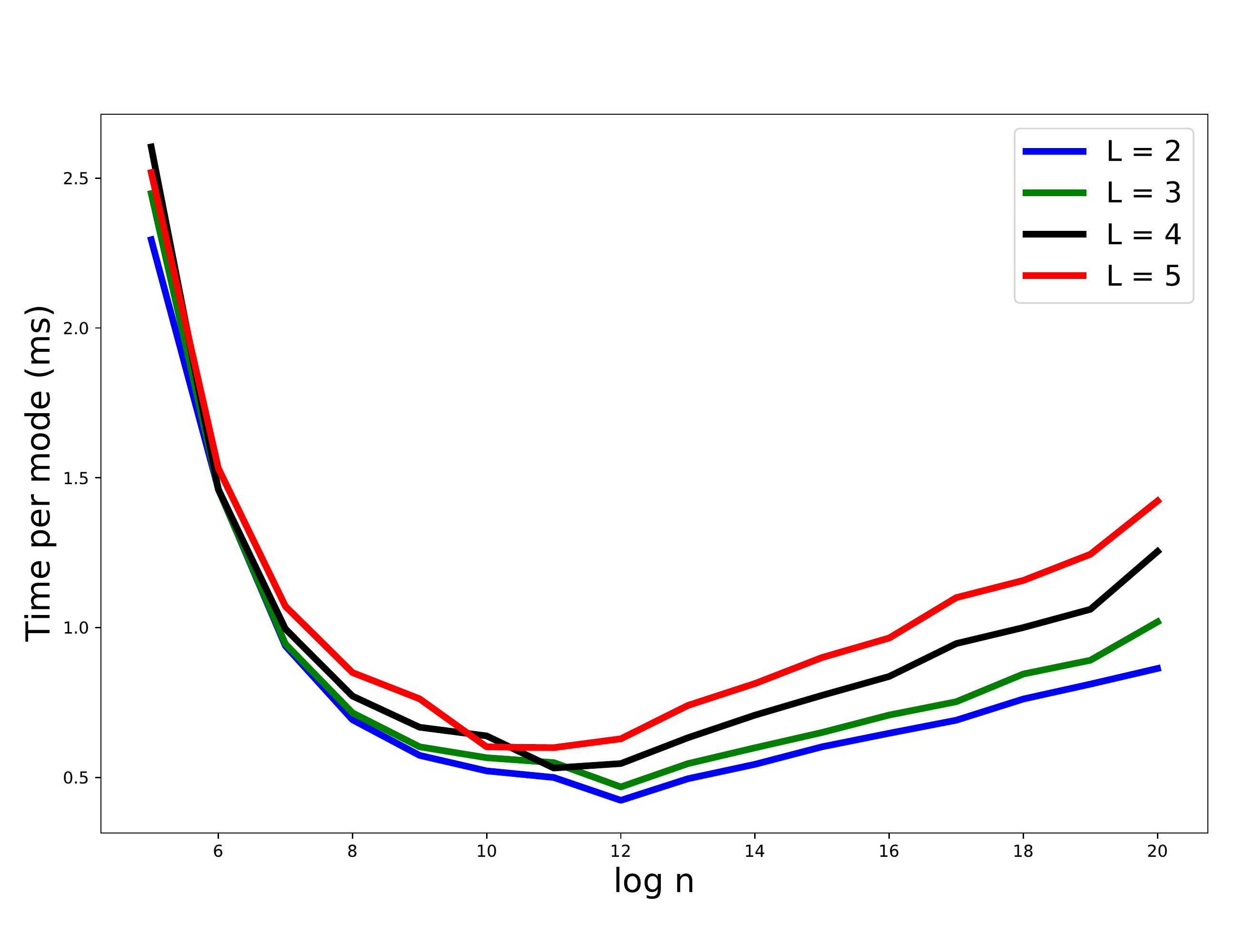}}
  \caption{The reconstruction error and the computing time per mode for different $L$. The reconstruction error of a holographic system is also shown for comparison. For $\log n<8$  the time to launch the kernel dominates the total runtime.}
  \label{fig:random_l_mode} 
\end{figure}

We solve for $x$ by first-order gradient-based optimization using the Adam optimizer implemented in TensorFlow~\cite{tensorflow2015-whitepaper, kingma2014adam}. MSE and computation time as a function of $n$ are summarized in Figure \ref{fig:random_l_mode}. In these simulations we independently draw the value of each mode in the input field from complex standard normal distribution $CN(0,1)$. Measurements are drawn from a Poisson distribution with $\bar{K}=10^4$ per mode. We repeated the simulation several times for each value of $n$ and averaged the results. As shown in the figure and expected from the results of the previous section, MSE falls as $L$ and $n$ increase. The slight rise in MSE at high dimension for $L=5$ is associated with quantization error in dividing $10^4$ photons into 100 measurements, increasing the number of photons per mode decreases the error. Figure \ref{fig:random_l_time} shows the computation time per mode, i.e., total runtime divided by the signal dimension. For the operating system and other hardware supports, we use Ubuntu 18.04, Intel Core i7-7700 @ 3.60GHz and 16 GB of RAM. The computer code used for this paper is available online~\cite{code}.



\begin{figure}[htbp]
  \centering
  \subfigure[Connections between modes]{
    \label{fig:connection_strategies} 
    \includegraphics[width=0.51\textwidth]{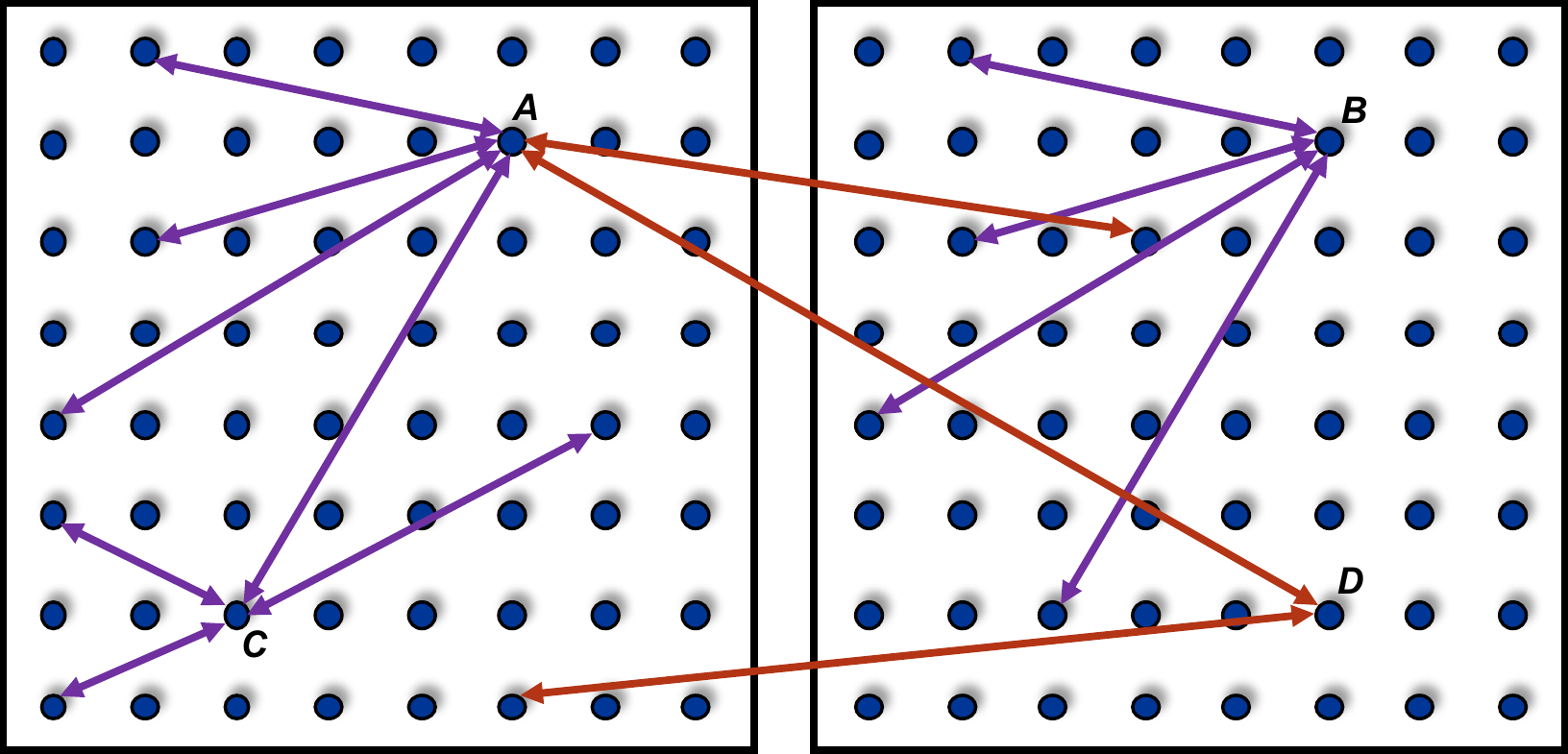}}
  \hspace{0.2in}
  \subfigure[The multiscale connection strategy]{
    \label{fig:multiscale1d} 
    \includegraphics[width=0.41\textwidth]{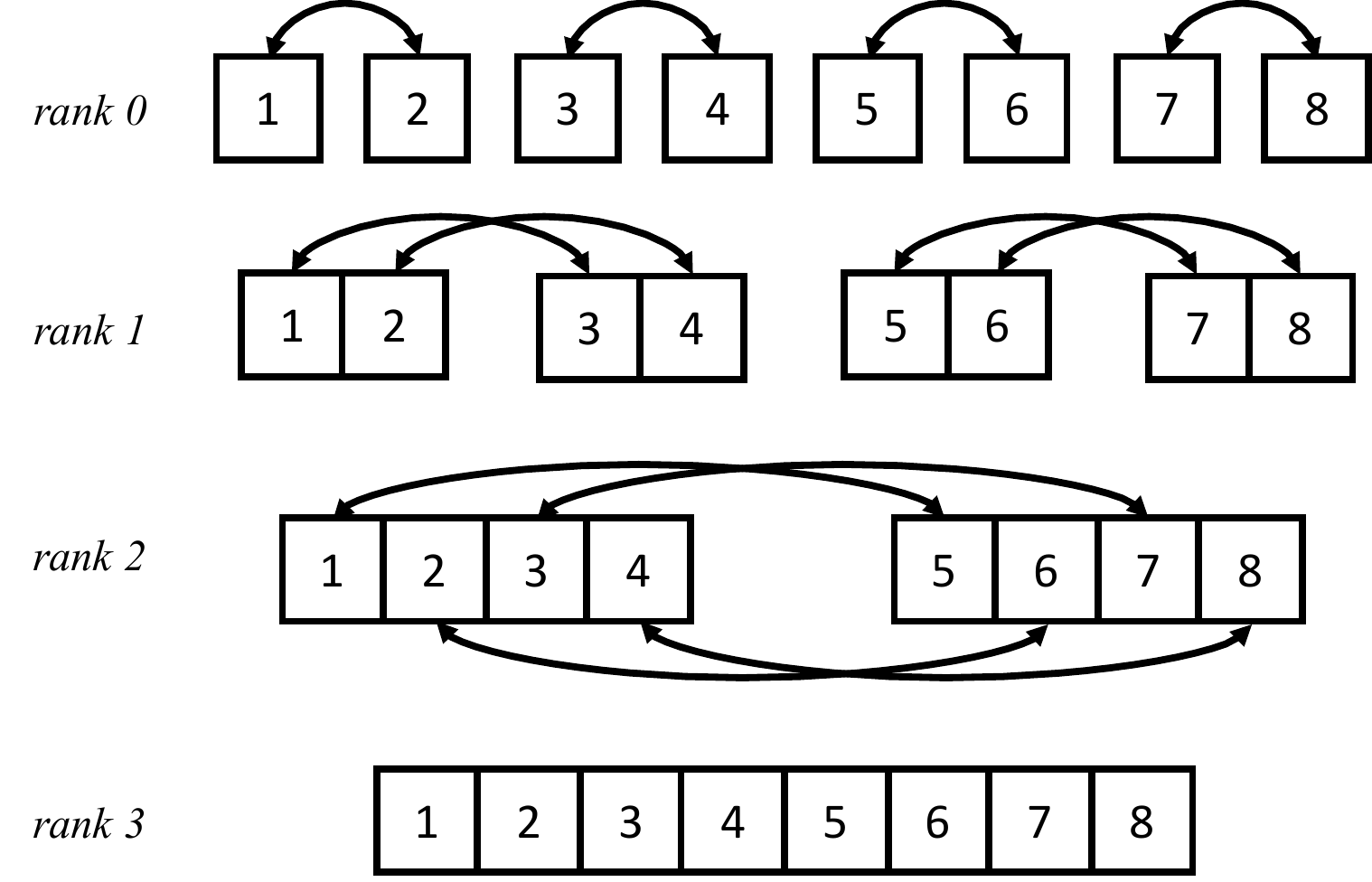}}
  \caption{Illustration of the multiscale system. (a) Connections between modes. The bounding boxes represents the {\it rank 0} blocks, the blue dots represent the mode to measure, and the arrows represent the connections between modes within (purple) and across (red) the block. (b) The multiscale connection strategy. The labelled squares represent {\it rank 0} blocks, the arrows represent the connections between blocks, and the blocks are put together when the relative phases between them are retrieved.}
  \label{fig:multiscale} 
\end{figure}

$n$ is ultimately limited by computing resources. For example, to process a $1024\times 1024$ image ($2^{20}$ modes) requires $>10^7$ measurements and an optimization engine with at least 10GB RAM. As $n$ increases additional challenges arise in interferometer design, physical implementation and numerical estimation. One approach to scalable design may be to build the interferometer network for a large field on top of blocks from smaller fields. This system is illustrated in Figure \ref{fig:multiscale}. The field {\bf x} is divided into several blocks, called {\it rank 0} blocks. Within each {\it rank 0} block, random connections and phase retrieval can be applied as described above. The two {\it rank 0} blocks are called connected when each mode in one block is randomly connected to one or more modes in another block. The relative phase between the two connected blocks can be computed from the cross-block connections. Figure \ref{fig:connection_strategies} shows an example of two connected {\it rank 0} blocks: each mode is connected to 4 modes (mode A, B and C) within the block and 2 modes across the block (mode A and D).


We find the global phase by successively building higher rank blocks. This process is illustrated in Figure \ref{fig:multiscale1d} with 8 {\it rank 0} blocks. First we build {\it rank 1} blocks by connecting pairs of neighboring {\it rank 0} blocks and applying proper phase shift to one of the block. Then each {\it rank 1} block consists of 2 {\it rank 0} blocks. Next we build {\it rank 1} blocks by pairing {\it rank 1} blocks. Each {\it rank 0} block is connected to the corresponding {\it rank 0} block in the other {\it rank 1} block, and the relative phase between the the {\it rank 1} blocks can be estimated from the all the cross-block connections. For example, in Figure \ref{fig:multiscale1d}, block 1 is connected to block 3, block 2 is connected to block 4, and the relative phase between the two {\it rank 1} blocks can be estimated by averaging the relative phases between the {\it rank 0} blocks. After phase shift, each {\it rank 2} block consists of 4 {\it rank 0} blocks. This process continues until all the relative phases are recovered.

\begin{figure}[htbp]
  \centering
  \includegraphics[width=0.7\textwidth]{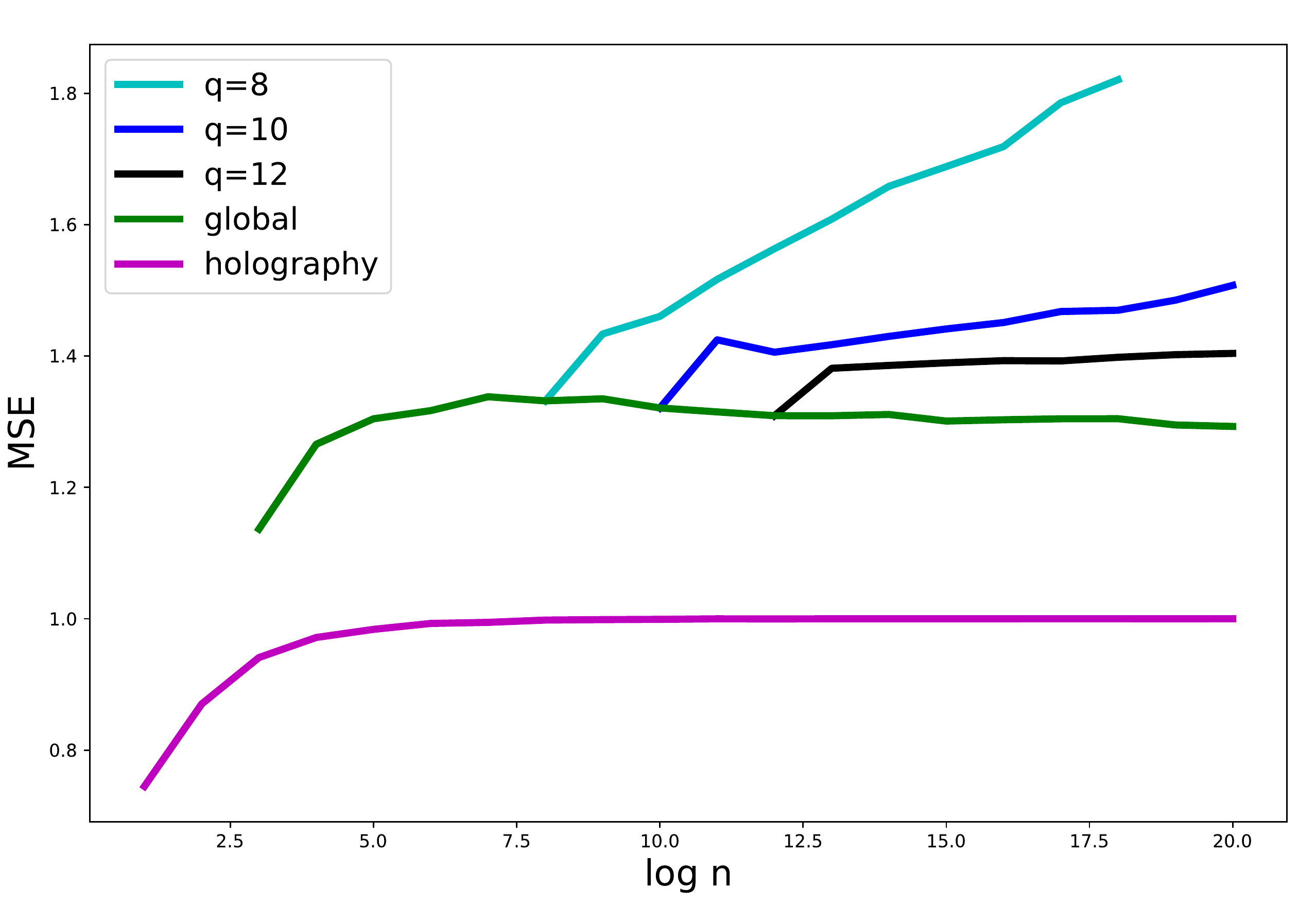}
  \caption{Simulation results on the multiscale system. Four different {\it rank 0} blocks ($q = 5,8,10,12$) are implemented, along with the global optimization and holography. In the global optimization method, when $n>12$, the random mapping is scaled up from $n=12$. }
  \label{fig:multiscale_result} 
\end{figure}

We test this multiscale system with simulations. We assume $L = 2$, $n = 2^k$ and $p = 2\log n$. There are $2^q$ modes in a {\it rank 0} block, each connected to another $2q$ modes in the same block, and each mode is connected to 2 more modes in another {\it rank 0} block when a higher rank block is to be built. Because the cross-block connections are only used to find the relative phases between blocks, only a short exposure time is necessary. In the simulation, only five percent of the signal energy is used for cross-block connections. The reconstruction error is shown in Figure \ref{fig:multiscale_result}. For $q=12$ reconstruction time is dominated by the intrablock phase retrieval, scaling to $n=2^{20}$ requires phase retrieval over 256 blocks of 4096 elements. The processing time per pixel remains approximately equal to the minimum associated with $n=2^{12}$ at the expense of a 0.7 dB increase in MSE. 

We have attempted to compare results presented here with previous phase retrieval studies, which have shown that $m\geq 4n-4$ measurements are sufficient to uniquely recover $x$~\cite{wang2017solving_raf, balan2006signal, conca2015algebraic}. The number of measurements made under our approach, $Ln\log n$, is much larger than in these studies, but the computational complexity of phase retrieval under our approach is much lower than previous work using  dense random matrices $A$, which require $O(n^2)$ operations per forward projection rather than the $O(n\log n)$ required here. Our computational complexity is similar to Fourier measurements, but simple Fourier transform sampling does not robustly achieve  quantum limited phase retrieval. Unfortunately, direct comparison between results is not possible because (1) previous methods fail to converge for a significant fraction of random inputs, rendering calculation of mean MSE impossible, and (2) higher computational complexity renders previous methods are impractical for $n>2^{12}$. In contrast, the proposed method here always recovers the signal. Because the proposed method uses sparse connections, built-in operators for sparse matrices in TensorFlow, or other platforms, can be used to speed computing and decrease memory, allowing for the reconstruction of megapixel fields.

\section*{Discussion}

This paper considers optimization of $\mathbf{A}$ for estimation of $\mathbf{x}$ from the intensities of $\mathbf{y}=\mathbf{A}\mathbf{x}$. In the presence of Poisson noise we prove that there exist $\mathbf{A}$ such that $\mathbf{x}$ can be estimated with quantum-limited precision. We further demonstrate in simulation that there exist practical measurement strategies that return $\mathbf{x}$ with precision near this limit. The implication of these results is that phase retrieval systems can match the precision of holographic measurement systems. We have not proven, but we here speculate that phase retrieval may be capable of exceeding holographic performance when other system parameters are considered. In particular, holography must assume that the reference and signal are mutually coherent whereas for phase retrieval the signal need only be self-coherent. 

We recognize that the interferometers proposed here will not be easily constructed, but we suggest that their construction forms a suitable challenge for emerging integrated photonics technologies. In practice, of course, the use of programmable interconnect networks in such devices could lead to measurement matrices significantly more advanced than those proposed here. Beyond issues of physical implementation, the results presented here raise more technological and theoretical questions than they answer. For example, the performance of the proposed system in analyzing partially coherent fields and natural image data remains to be explored. 

\bibliographystyle{unsrt}  
\bibliography{refs}





\end{document}